\documentclass[manuscript, screen]{acmart}
\AtBeginDocument{%
  }

\setcopyright{acmlicensed}
\copyrightyear{2025}
\acmYear{2025}
\acmConference{arXiv.org ’26}{ArXiv}{Online}
\usepackage{listings}
\usepackage{xcolor}
\definecolor{codegreen}{rgb}{0,0.6,0}
\definecolor{codegray}{rgb}{0.5,0.5,0.5}
\definecolor{codepurple}{rgb}{0.58,0,0.82}
\definecolor{backcolour}{rgb}{0.95,0.95,0.92}

\usepackage{tikz}
\usetikzlibrary{shapes.geometric, arrows}

\tikzstyle{startstop} = [rectangle, rounded corners, minimum width=3cm, minimum height=1cm, text centered, text width=3cm, draw=black, fill=orange!60]
\tikzstyle{process} = [rectangle, minimum width=3cm, minimum height=1cm, text centered, text width=3cm, draw=black, fill=orange!30]
\tikzstyle{decision} = [diamond, minimum width=3cm, minimum height=1cm, text centered, draw=black, fill=yellow!30]
\tikzstyle{arrow} = [thick,->,>=stealth]

\begin{document}

\title[Ground-Truth Depth in Vision Language Models]{Ground-Truth Depth in Vision Language Models: Spatial Context Understanding in Conversational AI for XR-Robotic Support in Emergency First Response.\\
\textcolor{white}{These Authors:}
\\
}

\author{Rodrigo Gutierrez Maquilon}
\email{rodrigo.gutierrez@ait.ac.at}
\orcid{0000-0002-6736-3418}
\affiliation{%
  \institution{AIT - Austrian Institute of Technology.}
  \department{Center for Technology Experience}
  \city{Vienna}
  \state{Vienna}
  \country{Austria}
}

\author{Marita Hueber}
\email{marita.hueber@ait.ac.at}
\affiliation{%
  \institution{AIT - Austrian Institute of Technology}
  \department{Center for Technology Experience}
  \city{Vienna}
  \state{Vienna}
  \country{Austria}
}

\author{Georg Regal}
\email{georg.regal@ait.ac.at}
\affiliation{%
  \institution{AIT - Austrian Institute of Technology}
  \department{Center for Technology Experience}
  \city{Vienna}
  \state{Vienna}
  \country{Austria}
}

\author{Manfred Tscheligi}
\email{manfred.tcheligi@ait.ac.at}
\affiliation{%
\institution{AIT - Austrian Institute of Technology}
\department{Center for Technology Experience}
 \city{Vienna}
  \state{Vienna}
  \country{Austria}
}

\renewcommand{\shortauthors}{Gutierrez Maquilon et al.}

\begin{abstract}

Large language models (LLMs) are increasingly used in emergency first response (EFR) applications to support situational awareness (SA) and decision-making, yet most operate on text or 2D imagery and offer little support for core EFR SA competencies like spatial reasoning. We address this gap by evaluating a prototype that fuses robot-mounted depth sensing and YOLO detection with a vision language model (VLM) capable of verbalizing metrically-grounded  distances of detected objects (e.g., the chair is 3.02 meters away). In a mixed-reality toxic-smoke scenario, participants estimated distances to a victim and an exit window under three conditions: video-only, depth-agnostic VLM, and depth-augmented VLM. Depth-augmentation improved objective accuracy and stability, e.g., the victim and window distance estimation error dropped, while raising situational awareness without increasing workload. Conversely, depth-agnostic assistance increased workload and slightly worsened accuracy. We contribute to human SA augmentation by demonstrating that metrically grounded, object-centric verbal information supports spatial reasoning in EFR and improves decision-relevant judgments under time pressure. 


\end{abstract}


\begin{teaserfigure}
  \includegraphics[width=\textwidth]{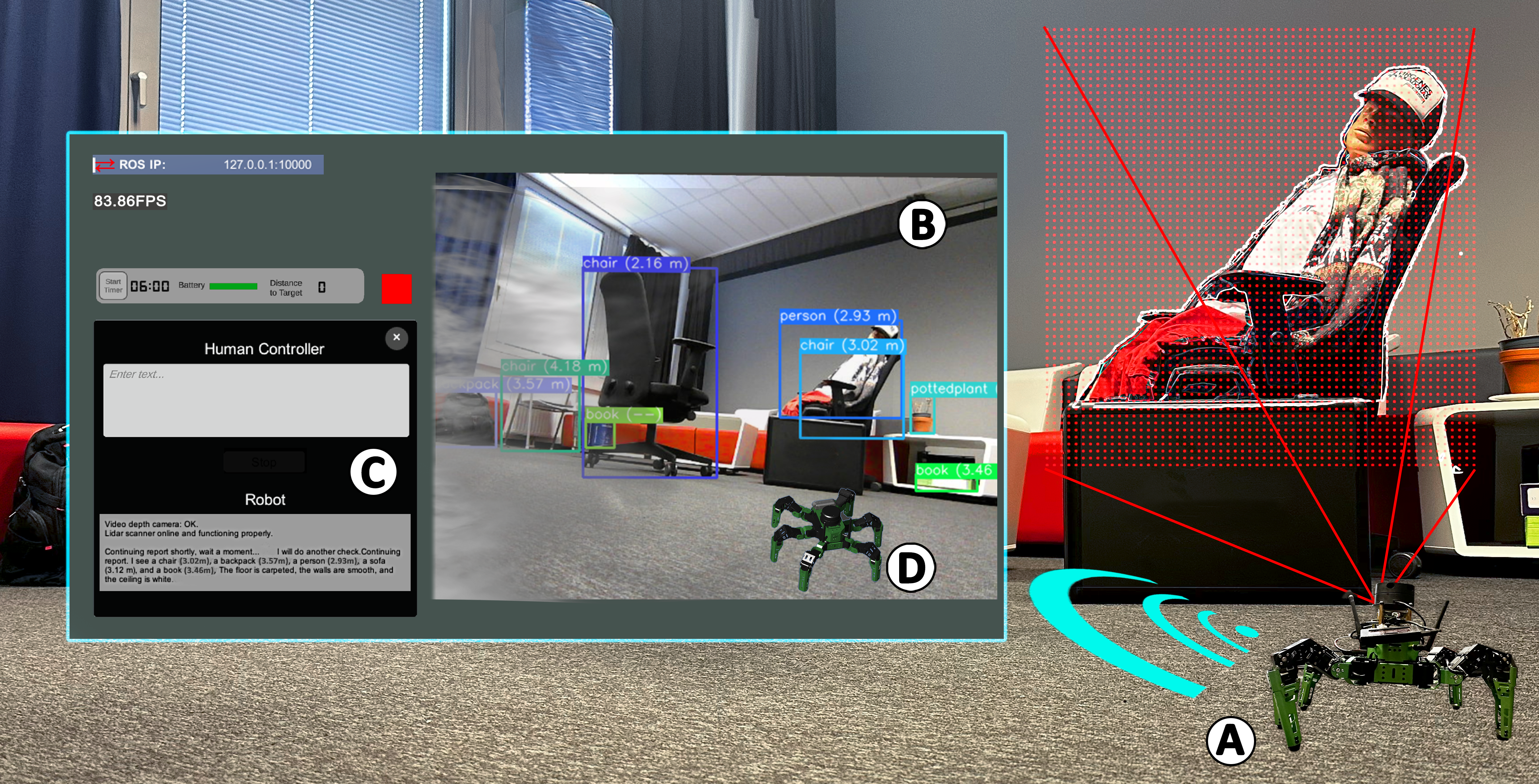}
  \caption{Mixed reality (MR) emergency response simulation of toxic smoke inside an office space. Depth measurements from the camera mounted on the robot (A) are shown in the labels of detected objects in MR view (B). Corresponding verbal feedback, e.g., "The person is approximately 2.93m away.", is shown on the chat window (C) and spoken through headphones. The robot also transmits its pose to enable 3D model tracking and visualization (D), e.g., behind obstacles. These audiovisual descriptions of the real-world environment are informed by ground truth depth.}
  \Description{A Unity 3D scene of a traffic accident where a virtual instructor provides feedback to the user about the treatment and communication with the virtual patient.}
  \label{fig:teaser}
\end{teaserfigure}


\maketitle

\section{Introduction}

Situational awareness (SA) is a core competence for emergency first response (EFR), especially in complex and time-critical scenarios such as mass-casualty incidents, structural collapses, or chemical, biological, radiological, nuclear and explosives (CBRNE) events \cite{Baetzner2022,Zechner2023,OBRIEN2020101634}. Beyond recognizing relevant objects and hazards, responders must continuously maintain a mental model of spatial relations: how far they are from a victim, whether an exit route is passable, or how close a robot can safely approach a suspected source of contamination \cite{doi:10.1518/001872095779049543, Spelke2007}. 

A fundamental aspect of SA, human distance estimation or spatial reasoning, in such settings is known to be unreliable. Studies in commonly used EFR environments, real-world, extended reality (XR) and robotics, show systematic biases depending on viewing angle, occlusions, stress level, and the presence of smoke, darkness, or protective equipment, with people typically under or overestimating distances by several tens of percent in the near field \cite{CreemRegehr2015,10.1145/1077399.1077403,10.1145/2543581.2543590,996536}. In XR, which encompasses virtual, augmented and mixed reality (VR, AR, MR), and in robotics, distortions from the field of view, head mounted display (HMD) or camera optics, and tracking errors can further degrade depth perception \cite{LoomisKnapp2003,10.1145/3106155,10.1145/2543581.2543590}. Training can partly compensate for these effects, but in practice emergency responders still need technological support to achieve reliable spatial judgments under pressure \cite{s23104849,Baetzner2022,Zechner2023}.

State-of-the-art vision language models (VLMs) promise a form of “spatial intelligence” by allowing users to ask free-form questions about images or video, such as “How far is the person from the door?” or “Is there anything between me and the fire extinguisher?” \cite{ashqar2024advancingobjectdetectiontransportation,liu2024m2se_vtts,jiao2024trainingfreeadaptiveempiricalinsights}. However, most current models are trained predominantly on 2D internet images and text, with little or no access to accurate 3D ground truth \cite{zha2025enablellm3dcapacity}. When prompted for distances, they implicitly infer depth from monocular cues learned during pre-training object size, texture gradients, shadows, or perspective but have never been calibrated against metric measurements for typical EFR scenes \cite{9009796,9711226}. Therefore, in most of these systems large language models (LLMs) operate on purely textual or 2D visual input and remain largely ignorant of the underlying 3D geometry that constrains safe behavior in the field \cite{chen2025spatialllmmultimodalitydataurban, ma2025spatialllmcompound3dinformeddesign,zha2025enablellm3dcapacity,fu2024scenellmextendinglanguagemodel}.

Recent evaluations of multimodal LLMs on 3D reasoning benchmarks and synthetic environments indicate that they struggle with precise spatial relationships, often confusing “near” and “far,” misordering objects along the depth axis, or giving inconsistent numeric estimates \cite{9878756,hong20233dllminjecting3dworld,ma2025spatialllmcompound3dinformeddesign,11086426}. Their distance judgments are often comparable to or worse than untrained human estimates, and small changes in camera viewpoint can lead to entirely different answers \cite{CreemRegehr2015,10.1145/2543581.2543590}. 

This gap is particularly problematic for EFR support: an apparently competent conversational assistant that nonetheless misjudges whether a victim is “within reach” or “behind the smoke” risks negatively impacting user trust or, in the worst case, encouraging unsafe decisions \cite{doi:10.1518/001872095779049543,chen2024integrationlargevisionlanguage}. As a result, current EFR applications use LLMs in combination with XR and robotics primarily for prompted communication and documentation summaries, while their potential to augment spatial reasoning to support decision-making in hazardous environments is still under explored \cite{doi:10.1177/02783649251351658,10160969,GutirrezMaquiln2024,Zechner2023}.

A promising way to overcome these limitations is to combine LLM-based scene understanding with ground-truth metric information obtained from depth sensors such as RGB-D cameras, structured-light or time-of-flight sensors, or LiDAR mounted on robots \cite{10160969,doi:10.1177/02783649251351658,Yang_Liu_Zhang_Pan_Guo_Li_Chen_Gao_Li_Guo_Zhang_2025}. Instead of asking a VLM to infer depth from 2D appearance alone, the system can measure the distance to objects directly in the sensor frame, then expose these values to the model as additional input or structured annotations \cite{hong20233dllminjecting3dworld,Wald_2020_CVPR,11086426}. In XR-robotic EFR settings, where robots already carry depth cameras for navigation and obstacle avoidance and HMDs provide real-time information overlays, distance measurements are readily available but rarely exploited in the human-AI interface \cite{JRC138914,10018826}. Integrating depth allows the assistant to provide responses such as “The victim is approximately 0.8 meters in front of the robot” or “The fire source is about 12 meters behind the patient and to the left” with centimeter-level accuracy, while still using the VLM to transform these data into concise, context-aware descriptions \cite{doi:10.1177/02783649251351658,10160969,fu2024scenellmextendinglanguagemodel}. 

We argue that this hybrid approach combining data-driven visual recognition with physically grounded distance measurements can turn current vision language models from depth-agnostic commentators into reliable spatial copilots for EFR SA augmentation in XR-robotic environments currently developing next-generation solutions\cite{s23104849, ZHU20211}.


Given this background, in this work we explore whether augmenting a VLM with ground-truth depth from a robot-mounted camera improves spatial context understanding for EFR tasks in mixed reality. Therefore, we formulate our central research question as follows: \\

\textbf{RQ: Do depth-augmented VLMs provide better spatial reasoning support to first responders than (a) relying on human distance estimation from a video feed alone, or (b) using a depth-agnostic VLM?}\\

We hypothesize that integrating metric depth into the VLM pipeline will:

\addtolength{\leftmargini}{-0.5cm}
\begin{itemize} 
\item\textbf{(H1) significantly reduce errors in distance estimations to task-relevant objects compared to human estimates and a baseline VLM.}

 \item \textbf{(H2) lead to better task performance and subjective confidence in a simulated EFR scenario where decisions depend on stand-off distance, accessibility, and proximity to hazards.}

\end{itemize} 

To test these hypothesis, we conducted a controlled experiment. Participants had to estimate the distance of specific objects with different levels of support in an MR EFR scenario where a ground robot monitored a smoke-filled office and transmitted the feed from its mounted depth camera to show detected objects with corresponding distances on the MR HMD. In this experiment we systematically compared human distance estimation from the video feed alone, with added support from a VLM, and with added support of a depth-augmented VLM.

In summary, this paper is positioned at the intersection of XR-based EFR scenarios, spatial multimodal LLMs, and robot-mediated perception \cite{doi:10.1177/02783649251351658,liu2024m2se_vtts,Zechner2023}. The remainder of the paper is structured as follows: We first review existing uses of LLMs in EFR operations and training, highlighting their current focus on communication rather than spatial support \cite{GutirrezMaquiln2024,Baetzner2022}. We then discuss human spatial context awareness and its limitations, motivate why distance estimation is a critical yet error-prone component of EFR decision-making, and outline the shortcomings of present-day VLMs when reasoning about 3D structure from 2D data \cite{10.1145/2543581.2543590,zha2025enablellm3dcapacity}. Building on this, we propose the integration of depth measurements into the LLM interaction loop as a way to provide accurate, explainable distance information to human operators \cite{Yang_Liu_Zhang_Pan_Guo_Li_Chen_Gao_Li_Guo_Zhang_2025,10160969}. Subsequently we evaluate the developed approach in an experimental study and explore its potential to enhance human situational awareness and decision-making for EFR. Finally we discuss the study and provide insights for future work in this domain highlighting the contribution of spatial reasoning augmentation to the community. 

\section{Related Work}
While research of virtual and mixed reality training for emergency first response (EFR) is producing useful applications, the integration of LLMs and spatially aware perception is still at an early stage \cite{Baetzner2022,Zechner2023,Uhl2024}. Existing XR simulations predominantly focus on procedural skills, team coordination, and communication training, whereas the role of multimodal LLMs remains largely confined to dialog support and scenario scripting \cite{GutirrezMaquiln2024,Baetzner2022}. In parallel, advances in spatial multimodal models and robot-mediated perception demonstrate that visual language architectures can be extended with 3D information from RGB-D cameras, LiDAR, or SLAM pipelines \cite{Yang_Liu_Zhang_Pan_Guo_Li_Chen_Gao_Li_Guo_Zhang_2025,10160969,doi:10.1177/02783649251351658}. However, these approaches are rarely connected to human-in-the-loop EFR scenarios in XR \cite{chen2024integrationlargevisionlanguage,JRC138914}. In this section we review: (1) current uses of LLMs in EFR operations and training, (2) human spatial context awareness and its limitations, and (3) VLMs for 3D reasoning and the potential of ground-truth depth integration.

\subsection{Existing uses of LLMs in EFR training and operations}
Recent work has begun to explore LLMs as conversational components in EFR support for live operations. In these prototypes, LLMs are used to summarize radio communication, annotate incident logs, or convert sensor metadata into more readable descriptions for the command center \cite{chen2024integrationlargevisionlanguage,zha2025enablellm3dcapacity}. Some systems explore question-answering interfaces over map-based situational pictures or structured incident databases, allowing officers to query past incidents, weather reports, or resource status \cite{JRC138914}. Others experiment with LLM-based copilots for unmanned ground vehicles (UGVs), where the model helps interpret camera images or generates natural-language status updates for the operator \cite{10018826,ashqar2024advancingobjectdetectiontransportation}. However, in most cases the spatial reasoning burden critical in EFR is carried by classical perception and mapping modules, which output symbolic summaries that are then verbalized by the LLM \cite{doi:10.1177/02783649251351658,10160969}.

EFR training applications enabled with LLMs follow a similar pattern and typically embed LLMs in virtual patients, virtual instructors, or mixed-reality communication training  scenarios \cite{GutirrezMaquiln2024,Baetzner2022}. In XR medical simulations, LLM-driven virtual patients can produce context-appropriate utterances enabling EFR personnel to practice history taking, reassurance, and information gathering in a safe environment \cite{GutirrezMaquiln2024,Baetzner2022}. Similarly, virtual instructors powered by LLMs have the potential to provide on-demand explanations of protocols, clarify affordances of virtual tools, and scaffold step-by-step procedures during simulated emergencies \cite{Uhl2024,Baetzner2022}. These systems have shown improvements in usability and perceived realism compared to purely scripted or menu-based interfaces, since the conversational agents can adapt their responses to user questions and to the evolving physiological state of the scenario \cite{Zechner2023}.

Despite this progress, the use of LLMs in EFR training remains predominantly communication-centric. Most implementations treat the model as a flexible dialog engine: it answers trainees’ questions, explains guidelines, or role-plays different stakeholders, but it does not have direct access to the embodied spatial state of the training environment \cite{chen2024integrationlargevisionlanguage,JRC138914}. Even when deployed in immersive XR settings, the LLM often receives only textual descriptions of events or high-level scenario labels rather than the live sensor data driving the simulation \cite{GutirrezMaquiln2024}. As a result, an AI instructor can talk about a dangerous leak, a blocked exit, or a trapped victim only to the extent that these facts are encoded in the prompt by the designer or a separate rule-based subsystem \cite{GutirrezMaquiln2024}.

In both operational and training settings, this architecture leads to a clear division of labor: low-level modules handle perception and 3D mapping, while the LLM sits at the top of the stack as a powerful but essentially 2D and text-focused commentator \cite{chen2024integrationlargevisionlanguage,fu2024scenellmextendinglanguagemodel}. It can infer some spatial relations from textual descriptions (for example, knowing that a victim "behind the vehicle" is not directly visible) but has no direct access to metric distance information or to the geometry of the scene as perceived through cameras and depth sensors \cite{Yang_Liu_Zhang_Pan_Guo_Li_Chen_Gao_Li_Guo_Zhang_2025}. Consequently, the current generation of LLM-enhanced EFR systems primarily augments communication, e.g., who says what, when, and how, rather than the spatial decision-making required to navigate hazardous environments, maintain safe stand-off distances, or prioritize reachable victims \cite{JRC138914,chen2024integrationlargevisionlanguage}. Bridging this gap requires bringing spatial context into the conversational loop, rather than relying on pre-encoded textual summaries \cite{doi:10.1177/02783649251351658,10160969}.

\subsection{Human spatial context awareness and its limitations}
Effective EFR performance depends heavily on situational awareness, and within it, spatial context awareness plays a central role \cite{doi:10.1518/001872095779049543,Baetzner2022}. Responders must quickly understand not only what is happening and to whom, but also where critical entities are located relative to themselves, to each other, and to environmental hazards \cite{Zechner2023}. Decisions such as approaching a victim, choosing a safe path, selecting a stand-off distance for a suspected CBRNE source, or positioning equipment in a confined space all depend on reliable distance judgments and relative positioning \cite{Spelke2007}. In XR-based training scenarios, these demands are mirrored: trainees need to interpret the virtual or mixed scene, estimate distances to victims, exits, and hazards, and make decisions under time pressure that rely on their internal spatial model \cite{Uhl2024,Zechner2023}.

Human ability to estimate distance is known to be systematically biased and context-dependent \cite{10.1145/2543581.2543590,CreemRegehr2015}. In real-world settings, people rely on a combination of monocular cues (e.g., relative size, texture gradients, motion parallax), binocular disparity, and proprioceptive feedback to judge how far away objects are \cite{LoomisKnapp2003,CuttingVishton1995}. These cues can be degraded or distorted in typical EFR conditions: smoke, darkness, debris, and protective gear all reduce visibility and constrain head movement; stress and cognitive load further diminish the precision of spatial judgments \cite{Baetzner2022,10.1145/1077399.1077403}. Empirical studies have shown that under such conditions people frequently underestimate or overestimate distances by a considerable margin, especially in the near to mid range where many tactical decisions are made \cite{10.1145/2543581.2543590,996536}. In tunnel, industrial, or urban search-and-rescue scenarios, even small misjudgements can translate into unsafe approaches to hazards or unnecessarily conservative stand-off distances that slow down rescue \cite{JRC138914}.

XR environments introduce additional sources of error. The optics of HMDs, their field of view, and their rendering of stereo cues can affect depth perception \cite{CreemRegehr2015,LoomisKnapp2003}. Calibration mismatches, tracking jitter, or latency between physical and virtual movement may disrupt the coupling between vestibular and visual cues \cite{10.1145/3106155}. While many users adapt over time, research on distance estimation in VR and MR consistently reports compression of perceived distances, particularly beyond a few meters \cite{10.1145/2543581.2543590,10.1145/1077399.1077403}. In training scenarios, this means that trainees may feel closer or farther from a victim or hazard than they actually are, depending on the display setup and scenario design \cite{Uhl2024}. If these distortions are not explicitly addressed, training may inadvertently reinforce biased spatial heuristics that carry over to field operations \cite{Baetzner2022}.

From a pedagogical perspective, this raises two challenges. First, trainers must be aware that human spatial context awareness is fallible and can differ between physical and XR environments \cite{CreemRegehr2015,10.1145/2543581.2543590}. Second, any digital assistant introduced into the training pipeline must not amplify these biases \cite{doi:10.1518/001872095779049543}. If a conversational AI simply echoes or loosely interprets the trainee’s own distance estimates, it risks reinforcing incorrect judgments. Conversely, if it contradicts human intuition without clear justification, it may be perceived as untrustworthy and be ignored \cite{doi:10.1518/001872095779049543}. For EFR, the stakes are high: distance is not just a geometric parameter but a proxy for risk, reachability, and intervention timing \cite{JRC138914}.

These limitations motivate the search for supportive technologies that can provide more reliable spatial information without overwhelming the responder \cite{JRC138914}. Range finders, laser pointers, and AR overlays are traditional solutions, but they often require additional manual effort or visual attention \cite{CreemRegehr2015}. Robot-mounted depth cameras already generate accurate metric information for navigation and obstacle avoidance, yet this information is rarely translated into human-understandable, context-aware feedback \cite{Yang_Liu_Zhang_Pan_Guo_Li_Chen_Gao_Li_Guo_Zhang_2025,10160969}. A spatially competent conversational agent that can verbalize distance information grounded in sensor data rather than human guesswork offers a promising route to improve spatial context awareness in both training and operations, while aligning with existing EFR practices of radio-based communication and verbal briefings \cite{doi:10.1177/02783649251351658}.

\subsection{Current Multimodal Spatial LLMs for 3D reasoning}

Vision language models extend LLMs with image or video input, enabling them to answer questions about what is visible in a scene, describe objects, and reason about simple spatial relations \cite{liu2024m2se_vtts,ashqar2024advancingobjectdetectiontransportation}. At first glance, this seems to provide exactly what EFR applications need: an operator could ask "How far is the victim from the door?" or "Is there any obstacle between the robot and the exit?" and receive a natural-language response \cite{jiao2024trainingfreeadaptiveempiricalinsights}. However, contemporary VLMs are typically trained on large corpora of 2D internet images paired with text, rather than on datasets with explicit 3D ground truth \cite{zha2025enablellm3dcapacity}. Their understanding of depth and spatial relationships is therefore implicit, relying on learned correlations between appearance and typical spatial configurations, not on calibrated metric information \cite{9711226,9009796}.

Approaches that rely solely on 2D data attempt to approximate 3D understanding via monocular depth prediction, geometric reasoning over image features, or multi-view consistency \cite{9711226,9009796}. In practice, these models can often distinguish qualitative relations such as "in front of", "behind", or "next to" in prototypical views, but they struggle with precise judgments, occlusions, and unusual camera angles \cite{9878756,9981261}. Evaluations on spatial reasoning benchmarks reveal frequent errors in ordering objects along the depth axis, confusion about which object is closer to the camera, and large variance when estimating distances in meters \cite{hong20233dllminjecting3dworld,11086426,ma2025spatialllmcompound3dinformeddesign}. Moreover, small changes in viewpoint or phrasing of the question can lead to inconsistent answers, undermining user trust \cite{zha2025enablellm3dcapacity}. For EFR use, such inconsistencies are particularly problematic: an assistant that alternately claims a victim is "within arm’s reach" and "several meters away" from nearly identical frames cannot be relied on for safety-critical decisions \cite{doi:10.1518/001872095779049543}.

To address these limitations, a growing body of work explores integrating explicit 3D information into multimodal LLM pipelines. Some methods fuse point clouds or voxelized 3D features with language, enabling the model to access geometric structure directly \cite{Yang_Liu_Zhang_Pan_Guo_Li_Chen_Gao_Li_Guo_Zhang_2025,11086426,hong20233dllminjecting3dworld}. Others combine SLAM-based maps, semantic segmentation, and object detection into spatial language maps that relate symbolic entities to positions in a global coordinate frame \cite{doi:10.1177/02783649251351658,10160969,fu2024scenellmextendinglanguagemodel}. These approaches demonstrate improved performance on tasks like 3D captioning, navigation instruction following, and object-goal navigation \cite{doi:10.1177/02783649251351658,10160969}. However, they often target autonomous agents or offline scene understanding rather than human-in-the-loop decision support, and their data requirements (dense 3D scans, carefully aligned multimodal datasets) can be difficult to satisfy in time-critical, hazardous EFR environments \cite{zha2025enablellm3dcapacity,9878756}.

More similar approaches to ours, investigate how depth sensors and object detectors can be combined to generate structured, distance-aware descriptions of a scene for robot navigation \cite{10160969}. In these pipelines, a 2D detector such as YOLO identifies task-relevant objects with bounding box category labels \cite{YOLOv8,SAPKOTA2026103575}. This information is then passed to an LLM, either as explicit annotations or embedded within template text, enabling the model to discuss contextual features without having to infer them from appearance alone \cite{Yang_Liu_Zhang_Pan_Guo_Li_Chen_Gao_Li_Guo_Zhang_2025,doi:10.1177/02783649251351658,SAPKOTA2026103575}. This hybrid approach respects the strengths and weaknesses of each component: classical perception handles detection and spatial relations, while the LLM focuses on explanation, contextualization, and dialog management \cite{fu2024scenellmextendinglanguagemodel,chen2025spatialllmmultimodalitydataurban}.

Our proposed system follows this hybrid design. Rather than attempting to train a fully 3D-aware VLM end-to-end, we integrate ground-truth depth from a robot-mounted camera into the LLM interaction loop at runtime. For each detected object, we compute its distance and relative position in the robot’s frame and expose these quantities as structured input to the conversational agent \cite{doi:10.1177/02783649251351658}. When a trainee or operator asks about the spatial layout, the model can respond with accurate, sensor-based distances (e.g., "The victim is approximately 0.8 meters in front of the robot and slightly to the right."), while still adapting its phrasing and level of detail to the user’s needs \cite{10160969}. This strategy sidesteps the scarcity of 3D training data by leveraging the mature ecosystem of depth sensors and object detectors already deployed on robots, and it aligns naturally with EFR workflows, where verbal communication is the primary channel for sharing spatial information \cite{Yang_Liu_Zhang_Pan_Guo_Li_Chen_Gao_Li_Guo_Zhang_2025,JRC138914}.

In summary, prior work on XR-based EFR training has successfully integrated LLMs for conversational support but largely ignores the underlying 3D geometry of the scene \cite{GutirrezMaquiln2024,Baetzner2022}. At the same time, research on spatial multimodal models and 3D-aware LLMs highlights both the potential and the current limitations of purely data-driven 3D reasoning \cite{zha2025enablellm3dcapacity,fu2024scenellmextendinglanguagemodel}. Our approach positions itself at the intersection of these strands by combining robot-mediated ground-truth depth with VLMs to provide accurate, distance-aware descriptions in mixed reality \cite{Yang_Liu_Zhang_Pan_Guo_Li_Chen_Gao_Li_Guo_Zhang_2025,10160969,doi:10.1177/02783649251351658,SAPKOTA2026103575}. This enables a new form of spatial conversational support that complements human perception and addresses specific shortcomings of current EFR training and support systems \cite{JRC138914}.

\section{Method}

In this section we describe the controlled exploratory study we conducted to address our research question.

\subsection{Mixed Reality Simulation}
An MR emergency response simulation was presented to the participants through an HMD. The simulation consisted of virtual toxic smoke inside a real office space and a real ground robot with a mounted depth camera sent for inspection. A manikin inside the office was used as a simulated human victim going unconscious and unable to walk. Participants outside the closed office could only see the camera feed of the robot through the HMD. Participants must determine the distance of a window and the victim from the robot for assessment and further course of action (see Fig.1).

\subsection{Aparatus and VLM Interaction}
The meta Quest Pro HMD was used to show the robot's camera feed. Unity 6.2.7f2 was used on a Windows PC with 13th gen Intel i9-13900K (3.00 GHz) CPU  with 64GB of RAM and an Nvidia RTX 4080 GPU to build and run an application that subscribed to the robot's depth camera feed and distance measurements using ROS-Unity integration. The application further integrated the depth values with the labels of detected objects from yolov8x.onnx. The VLM qwen2.5vl:32b was prompted to describe the snapshots taken when a participant said the keyword "report" followed by the question "How far is the window?". The Unity application also subscribed to the pose of the robot to overlay its 3D model on it for real-time tracking and display on the HMD outside the office space. Finally, the application also generated the virtual smoke. The robot was a Jethexa by Hiwonder powered by the Jetson Nano micro computer with a mounted ORBBEC Dabai DCW depth camera transmitting RGB, depth and IR topics (see Fig.2).

\begin{figure}[h]
  \includegraphics[width=1\textwidth]{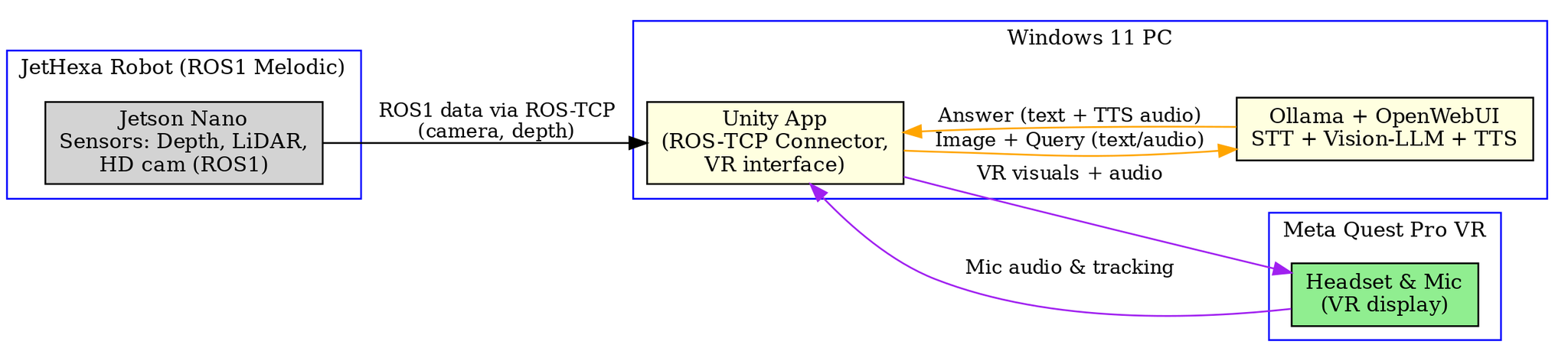}
  \caption{Integration diagram of the Jethexa robot, Unity application, vision language model and mixed reality head mounted display.}
  \Description{An office space with a manikin, chairs, backpack and table. Labels of detected objects with their distance from the camera are overlayed on each object.}
  \label{fig:teaser}
\end{figure}

\subsection{Study Design}
The simulation was presented in 3 different conditions: a baseline condition (C1) presented to all participants and two treatment conditions (C2 and C3) assigned to two groups of different participants. The conditions were defined as follows: 

 \begin{itemize}
 \item Condition1: Participants relied only on the RGB camera feed in the HMD and on their human ability to determine the distance of a window inside the office and the manikin representing a human victim in the MR simulation.

 \item Condition2: Participants additionally relied on the verbal interaction with qwen2.5vl:32b to get a description of the scenario without the integration with the depth camera measurements and determine the distance of the window and the manikin. 

 \item Condition3: Participants additionally relied on the verbal interaction with qwen2.5vl:32b to get a description of the scenario supported by the integration of depth camera measurement and YOLOv8x object detection and determine the distance of the window and the manikin. (See Fig.3).
 \end{itemize} 

 \begin{figure}[h]
  \includegraphics[width=0.75\textwidth]{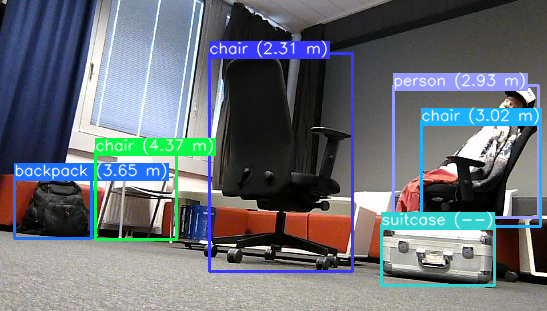}
  \caption{Condition 3. Vision language model qwen2.5vl:32b capture of depth camera feed with distance measurements integrated in the labels of YOLOv8x detected objects.}
  \Description{An office space with a manikin, chairs, backpack and table. Labels of detected objects with their distance from the camera are overlayed on each object.}
  \label{fig:teaser}
\end{figure}

\subsection{Participants}
Sixteen participants (9 male, 7 female) took part in the study. Their ages ranged from 21 to 66 with a mean age of 48 years (SD = 9.13) and 6 male participants reported different levels of familiarity with emergency response ranging from firefighter to volunteer paramedic and first aid course attendees.

\subsection{Measurements}
After each condition, participants completed standardized questionnaires to assess situational awareness, workload, interaction quality, and usability. All questionnaires were presented digitally. 

\subsubsection{Situational Awareness}
Situational awareness was measured using the Situational Awareness Rating Technique (SART). This instrument assesses how well participants perceive, understand, and anticipate elements of a dynamic situation. The ten SART items measure perceived attentional demand, attentional supply, and understanding of the situation on seven-point scales ranging from “very low” to “very high.” Higher scores indicate greater perceived situational awareness \cite{SART}.

\subsubsection{Perceived Workload}
Workload was assessed using the NASA Task Load Index (NASA-TLX). This well-established measure captures six dimensions of subjective workload: mental demand, physical demand, temporal demand, effort, performance, and frustration.In this study, the original scales were transformed into a seven-point rating format, and participants evaluated each dimension accordingly. The overall workload score was then calculated as the average across all six subscales \cite{NASA_TLX}.

\subsubsection{Voice Interaction}
The Subjective Assessment of Speech System Interfaces (SASSI) was used to evaluate the perceived quality of the LLM’s speech interaction. The questionnaire measures how users perceive aspects such as system accuracy, reliability, cognitive demand, and annoyance. Each statement is rated from 1 (“strongly disagree”) to 7 (“strongly agree”), with higher mean scores indicating more positive evaluations of the system’s conversational qualities and interaction flow \cite{SASSI}.

\subsubsection{Perceived Usability}
Perceived usefulness was assessed using a single item from the Usability Metric for User Experience (UMUX-Lite). Participants rated the statement “This system’s capabilities meet my requirements” on a seven-point Likert scale ranging from “Strongly disagree” to “Strongly agree.” Higher scores indicate that the system was perceived as more useful and capable of fulfilling the task requirements \cite{UMUX}.

\subsubsection{Additional Questions}
Finally, participants responded to exploratory questions about their confidence in estimating spatial distances. After each condition, they rated how confident they were that their reported distances to two reference points, namely the person and the window handle, were within ±10 cm of the true values. This confidence measure served as a subjective proxy for perceived spatial accuracy additional to the calculations of objective error data. In addition, participants were invited to provide short written comments describing any perceived differences in support quality, ease of interaction, and overall experience between the conditions.

\subsection{Procedure}
At the beginning of the study, participants were welcomed individually and provided with a brief overview of the study purpose and structure. After reading an information sheet, each participant signed an informed consent form. Participants were then introduced to the experimental setup, which included the main user interface displaying a live video feed and, depending on the condition, additional multimodal feedback components such as voice, text, or depth cues.

Before the first condition began, a short demonstration phase familiarized participants with the interaction process and the system’s response behavior. This tutorial ensured that all participants understood how to operate the interface, interpret its feedback, and carry out the spatial estimation task.

Following the demonstration, each participant began with the baseline condition (C1), in which they viewed the video feed and were asked to estimate the distances from the camera to two reference points in the scene: a person (3.22 meters away) and a window handle (4.45 meters away). After completing this condition, participants filled out the post-condition questionnaires.

After completing the baseline surveys, participants proceeded to their assigned second condition. Even-numbered participants continued with the voice/text-based LLM condition (C2), while odd-numbered participants continued with the depth-augmented LLM condition (C3). The same estimation task was repeated, followed by the corresponding set of questionnaires.

Upon completion of the final survey, participants were thanked for their contribution, given the opportunity to share additional qualitative feedback, and received a small financial compensation for their time. The full study lasted approximately 30 to 40 minutes per participant.

\section{Results}
The following section presents the quantitative findings of the study. Descriptive and inferential statistics were used to compare participants’ responses across the two experimental transitions (C1→C2 and C1→C3). In addition, correlational analyses were conducted to explore how these measures interrelated and to identify patterns that explain how multimodal and depth-augmented feedback influence cognitive and affective aspects of task performance.

\subsection {Distance Estimations}

\begin{figure}[h]
  \includegraphics[width=0.75\textwidth]{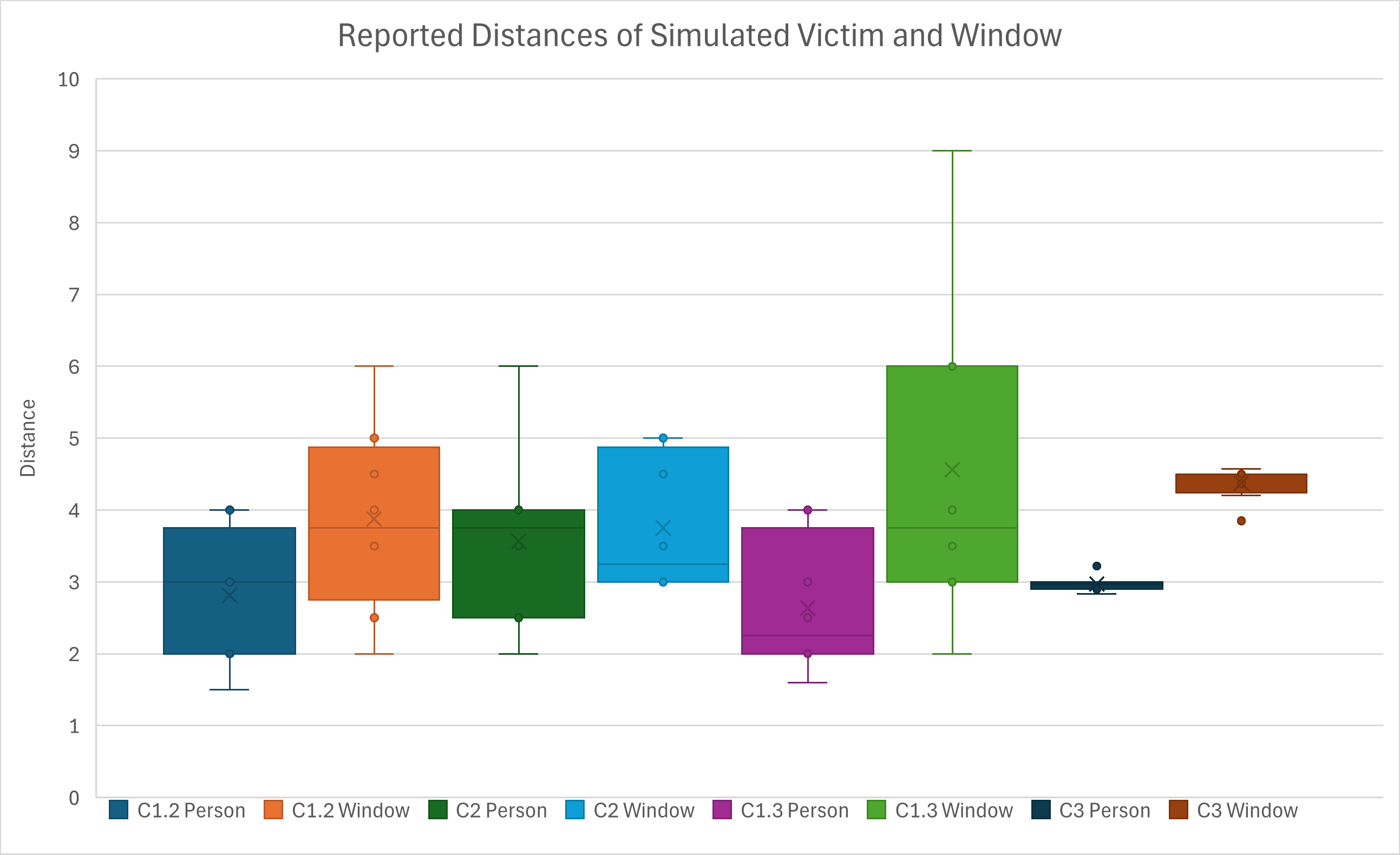}
  \caption{C1.2 and C1.3 = video-only baseline; C2 = VLM support; C3 = depth-augmented VLM support. }
  \Description{An office space with a manikin, chairs, backpack and table. Labels of detected objects with their distance from the camera are overlayed on each object.}
  \label{fig:teaser}
\end{figure}

For the person (true = 3.22m), mean estimates moved from 2.64m (C1) to 2.97m (C3), shrinking the absolute error from 0.58m to 0.25m and the SD from 0.94 to 0.12. For the window handle (true = 4.45m), mean estimates moved from 4.56m (C1) to 4.37m (C3), with absolute error dropping from 0.11m to 0.08m and SD from 2.29 to 0.24. In contrast, the depth-agnostic LLM (C2) slightly worsened accuracy relative to C1 (e.g., person: 2.81m → 2.56m; window: 3.88m → 3.75m). 

\subsection{Overview of Descriptive Trends Across Interaction Conditions}

\begin{table}[h]
\centering
\caption{Descriptive statistics (means and standard deviations) for situational awareness , perceived workload, voice interaction quality, perceived usability and confidence ratings across the three conditions.}
\label{tab:descriptive}
\begin{tabular}{lccc}
\toprule
\textbf{Measure} & \textbf{C1 Mean (SD)} & \textbf{C2 Mean (SD)} & \textbf{C3 Mean (SD)} \\
\midrule
NASA-TLX (Workload)       & 2.56 (1.16) & 3.30 (1.26) & 3.29 (1.74) \\
SART (Awareness)          & 3.70 (1.00) & 4.19 (0.91) & \textbf{4.74 (0.88)} \\
SASSI (Voice Interaction)  & 4.21 (0.86) & 4.41 (0.32) & \textbf{4.48 (1.12)} \\
UMUX (Usefulness Item)     & 4.86 (1.46) & 4.11 (1.54) & 5.00 (0.89) \\
Confidence (Distance Est.) & 3.89 (2.20) & 4.11 (2.42) & 4.57 (1.90) \\
\bottomrule
\end{tabular}

\vspace{0.5em}
\footnotesize \textit{Note.} Higher scores indicate greater workload, awareness, usability, or confidence on 7-point Likert scales. 
\end{table}

Situational Awareness Rating Technique (SART)
As shown in table \ref{tab:descriptive}, both LLM-supported conditions (C2 and C3) produced higher mean SART scores than the baseline, indicating that participants experienced greater situational awareness when assisted by LLM-based feedback. The depth-augmented condition (C3) achieved the largest improvement in SART (M = 4.74, SD = 0.88), accompanied by a workload level comparable to that of the baseline (M = 3.29 vs. 2.56). In contrast, the voice/text-only condition (C2) also enhanced awareness (M = 4.19), but this improvement coincided with a moderate rise in perceived workload (NASA-TLX M = 3.30).

Participants’ evaluations of voice interaction quality, as measured by SASSI, remained high and stable across all conditions, reflecting generally positive impressions of the speech interface. Likewise, perceived usability, assessed with the single-item UMUX statement (“This system’s capabilities meet my requirements”), was highest in the depth-augmented condition, suggesting that participants found the addition of spatial cues most aligned with their task needs. Participants reported greater confidence that their estimated distances to the person (3.22 m) and the window handle (4.45 m) were within ±10 cm of the actual values in both LLM-assisted conditions, with the strongest increase occurring under depth-augmented support (C3).

\subsection{Statistical Comparison of Baseline and LLM-Assisted Conditions}
To examine changes between the baseline and each assisted condition, Wilcoxon signed-rank tests were conducted for all dependent measures as illustrated in table \ref{tab:wilcoxon}.

\begin{table}[h]
\centering
\caption{Wilcoxon signed-rank tests comparing baseline (C1: video-only) with LLM-assisted conditions (C2: voice/text; C3: depth-augmented) across all dependent measures. Reported are the Wilcoxon W statistic and two-sided p-values.}
\label{tab:wilcoxon}
\begin{tabular}{lcccc}
\toprule
\textbf{Measure} & \textbf{W (C1$\rightarrow$C2)} & \textbf{p (C1$\rightarrow$C2)} & \textbf{W (C1$\rightarrow$C3)} & \textbf{p (C1$\rightarrow$C3)} \\
\midrule
Confidence        & 12.0 & .734 & 7.5 & .375 \\
NASA-TLX          &  5.0 & \textbf{.039} & 6.0 & .686 \\
SART              &  3.0 & \textbf{.035} & 0.0 & \textbf{.016} \\
SASSI             & 10.0 & .164 & 9.0 & .844 \\
UMUX (Lite item)  &  1.0 & .285 & 0.0 & .066 \\
\bottomrule
\end{tabular}

\vspace{0.5em}
\footnotesize \textit{Note.} For C1$\rightarrow$C2 comparisons, $n=9$ paired observations; for C1$\rightarrow$C3, $n=7$. W is the Wilcoxon signed-rank statistic; p-values are two-sided. 
\end{table}

The analysis revealed distinct patterns across the two transitions. From C1 → C2, both perceived workload (NASA-TLX; p = .039) and situational awareness (SART; p = .035) increased significantly. This suggests that while the voice/text-based LLM enhanced participants’ understanding of the spatial scene, it also demanded greater cognitive effort, likely due to the need to process and integrate auditory and textual information alongside the video feed.

In contrast, the C1 → C3 comparison showed a significant increase in situational awareness (SART; p = .016) without a corresponding rise in workload (NASA-TLX; p = .686). This pattern indicates that the addition of depth cues enabled participants to perceive and interpret spatial relationships more intuitively, reducing the cognitive demand associated with maintaining situational awareness.

Although SASSI scores did not differ significantly between conditions, their consistently high values across conditions indicate stable perceptions of system reliability and voice quality. The UMUX single-item ratings showed a trend toward higher perceived usefulness in the depth-augmented condition (p = .066), consistent with the descriptive pattern observed in Table \ref{tab:descriptive}. Confidence ratings also increased numerically in both groups, suggesting improved subjective accuracy in distance estimation, though these changes were not statistically significant.

\subsection{Interrelations Between Workload, Awareness, Usability, and Confidence}

\begin{table}[h]
\centering
\caption{Spearman correlation matrix for participants who completed C1$\rightarrow$C2, showing relationships among confidence, workload (NASA-TLX), situational awareness (SART), and voice interaction quality (SASSI) across both conditions.}
\label{tab:corr_c1c2}
\scriptsize
\begin{tabular}{lcccccccc}
\toprule
 & \textbf{C1\_CONF} & \textbf{C1\_NASA} & \textbf{C1\_SART} & \textbf{C1\_SASSI} & \textbf{C2\_CONF} & \textbf{C2\_NASA} & \textbf{C2\_SART} & \textbf{C2\_SASSI} \\
\midrule
\textbf{C1\_CONF}  & 1.00 & -0.32 &  0.22 &  0.79 &  0.36 & -0.37 &  0.34 &  0.78 \\
\textbf{C1\_NASA}  &-0.32 &  1.00 &  0.57 & -0.01 &  0.19 &  0.77 &  0.34 & -0.36 \\
\textbf{C1\_SART}  & 0.22 &  0.57 &  1.00 &  0.14 &  0.69 &  0.61 &  0.82 &  0.08 \\
\textbf{C1\_SASSI} & 0.79 & -0.01 &  0.14 &  1.00 &  0.10 & -0.41 &  0.27 &  0.85 \\
\textbf{C2\_CONF}  & 0.36 &  0.19 &  0.69 &  0.10 &  1.00 &  0.42 &  0.89 &  0.08 \\
\textbf{C2\_NASA}  &-0.37 &  0.77 &  0.61 & -0.41 &  0.42 &  1.00 &  0.44 & -0.61 \\
\textbf{C2\_SART}  & 0.34 &  0.34 &  0.82 &  0.27 &  0.89 &  0.44 &  1.00 &  0.27 \\
\textbf{C2\_SASSI} & 0.78 & -0.36 &  0.08 &  0.85 &  0.08 & -0.61 &  0.27 &  1.00 \\
\bottomrule
\end{tabular}

\vspace{0.5em}
\footnotesize \textit{Note.} $n=9$. Values are Spearman’s $\rho$ (bounded in $[-1,1]$). Larger positive values indicate stronger positive association; negative values indicate inverse association.
\end{table}

\begin{table}[h]
\centering
\caption{Spearman correlation matrix for participants who completed C1$\rightarrow$C3, showing relationships among confidence, workload (NASA-TLX), situational awareness (SART), and voice interaction quality (SASSI) across both conditions.}
\label{tab:corr_c1c3}
\scriptsize
\begin{tabular}{lcccccccc}
\toprule
 & \textbf{C1\_CONF} & \textbf{C1\_NASA} & \textbf{C1\_SART} & \textbf{C1\_SASSI} & \textbf{C3\_CONF} & \textbf{C3\_NASA} & \textbf{C3\_SART} & \textbf{C3\_SASSI} \\
\midrule
\textbf{C1\_CONF}  & 1.00 & -0.32 & -0.22 &  0.75 &  0.22 & -0.06 & -0.55 &  0.41 \\
\textbf{C1\_NASA}  &-0.32 &  1.00 &  0.38 & -0.49 & -0.26 &  0.83 &  0.54 & -0.49 \\
\textbf{C1\_SART}  &-0.22 &  0.38 &  1.00 & -0.12 & -0.27 & -0.06 &  0.70 & -0.17 \\
\textbf{C1\_SASSI} & 0.75 & -0.49 & -0.12 &  1.00 & -0.18 & -0.14 & -0.49 &  0.71 \\
\textbf{C3\_CONF}  & 0.22 & -0.26 & -0.27 & -0.18 &  1.00 & -0.44 & -0.62 & -0.62 \\
\textbf{C3\_NASA}  &-0.06 &  0.83 & -0.06 & -0.14 & -0.44 &  1.00 &  0.26 & -0.09 \\
\textbf{C3\_SART}  &-0.55 &  0.54 &  0.70 & -0.49 & -0.62 &  0.26 &  1.00 & -0.03 \\
\textbf{C3\_SASSI} & 0.41 & -0.49 & -0.17 &  0.71 & -0.62 & -0.09 & -0.03 &  1.00 \\
\bottomrule
\end{tabular}

\vspace{0.5em}
\footnotesize \textit{Note.} $n=7$. Values are Spearman’s $\rho$.
\end{table}

To further explore interrelations among workload, situational awareness, usability, and confidence, Spearman’s rank correlations were calculated separately for participants who experienced C1 → C2 and C1 → C3 as shown in the tables \ref{tab:corr_c1c2} and \ref{tab:corr_c1c3}

In the C1 → C2 subset, confidence was strongly and positively correlated with both conversational quality (SASSI) and situational awareness (SART), indicating that higher-quality speech interaction and greater environmental understanding co-occurred with stronger confidence in distance estimation. Workload (NASA-TLX) also correlated positively with awareness, suggesting that participants expended additional cognitive effort to achieve higher situational understanding when relying on voice and text feedback.

In contrast, the C1 → C3 subset revealed a different pattern. Confidence correlated negatively with both workload and situational awareness, implying that participants who felt more confident in their spatial judgments experienced lower mental demand, which is consistent with more efficient perceptual processing given the VLM support of depth information. The positive correlation of SART scores across conditions suggests that individual awareness tendencies remained stable, while the strong correlation of SASSI ratings indicates that perceptions of voice interaction quality were consistent even with the addition of depth cues.

\section{Discussion}

\subsection{Summary of findings relative to the research question}

In this study, we explored the question: \emph{“Do depth-augmented VLMs provide better spatial reasoning support to first responders than (a) relying on human distance estimation from a video feed alone, or (b) using a depth-agnostic VLM?”} 

Our pattern of results, significantly higher situational awareness without added workload under depth augmentation, together with objectively tighter and more accurate distance estimates, directly addresses limitations reported for communication-centric EFR assistants that lack embodied spatial access \cite{GutirrezMaquiln2024,Baetzner2022,chen2024integrationlargevisionlanguage,JRC138914}. In contrast to systems where classical perception produces symbolic summaries that are merely verbalized by an LLM  (e.g., “in between the sofa and TV”) \cite{doi:10.1177/02783649251351658,10160969}, our metrically grounded responses reduce the gap between 3D sensing and conversational guidance, aligning with recent calls to bring spatial context into the dialogue loop \cite{Yang_Liu_Zhang_Pan_Guo_Li_Chen_Gao_Li_Guo_Zhang_2025,JRC138914}. In summary, these improvements map onto core EFR spatial reasoning demands (stand-off, reachability) emphasized in XR training literature \cite{Zechner2023,Uhl2024}.

\subsection{Hypothesis H1: distance estimation accuracy}

H1 stated that integrating metric depth would \emph{“significantly reduce errors in distance estimations to task-relevant objects compared to human estimates and a baseline VLM.”} The observed accuracy improvements and variance decrease under depth augmentation are consistent with evidence that humans systematically misestimate distances, especially in visually degraded or XR settings due to cue conflicts and display factors \cite{10.1145/2543581.2543590,CreemRegehr2015,LoomisKnapp2003,10.1145/1077399.1077403}. They also address known weaknesses of depth-agnostic VLMs, which largely learn spatial regularities from 2D data and show instability on 3D reasoning benchmarks \cite{zha2025enablellm3dcapacity,9878756,9981261,hong20233dllminjecting3dworld,ma2025spatialllmcompound3dinformeddesign}. By injecting measured distances on command, our approach supports hybrid pipelines used in robotics detections with maps or depth sensors \cite{doi:10.1177/02783649251351658,10160969}. In short, our H1 test shows how 3D sensor information can rectify the “worse-than-human” distance judgments documented for monocular or text-only reasoning \cite{9711226,9009796,zha2025enablellm3dcapacity}. 

\subsection{Hypothesis H2: task performance and subjective confidence}

H2 proposed that depth integration would \emph{“lead to better task performance and subjective confidence in a simulated EFR scenario where decisions depend on stand-off distance, accessibility, and proximity to hazards.”} 

The combination of higher awareness (without extra workload), improved perceived usefulness, and better objective distance estimations under depth augmentation converges with human-factors theory: situation awareness increases when information is presented in task-relevant frames and with appropriate precision \cite{doi:10.1518/001872095779049543}. These outcomes extend prior XR EFR training systems where LLMs improve communication realism but remain spatially blind \cite{GutirrezMaquiln2024,Baetzner2022} by demonstrating that metrically grounded language can augment decision-making critical to EFR operations, which are misaligned with current 2D-only VLMs \cite{JRC138914,chen2024integrationlargevisionlanguage}. Although we did not yet include time-critical operational metrics, our results mirror studies reporting that spatial language is tied to robot maps or scene graphs \cite{doi:10.1177/02783649251351658,10160969}, suggesting a clear pathway from subjective and behavioral proxies to mission-level measures in future studies.

\subsection{Why depth helps: cognitive efficiency and representational alignment}

Depth augmentation appears to reduce the cognitive “translation” cost that arises when operators must reconcile qualitative utterances with ambiguous 2D cues that reflected in higher workload for depth-agnostic assistance. This interpretation resonates with XR perception research showing that display optics, FOV, and tracking can compress perceived space and elevate uncertainty \cite{10.1145/2543581.2543590,LoomisKnapp2003,10.1145/3106155}. By directly anchoring utterances to sensor-measured distances, our "depth-aware VLM" targeted spatial language maps and 3D-aware VLM pipelines, bringing language onto the coordinates that robots already use to act \cite{doi:10.1177/02783649251351658,10160969,fu2024scenellmextendinglanguagemodel}. This closes the gap identified in deployments where LLMs are used as commentators rather than spatial partners \cite{chen2024integrationlargevisionlanguage,JRC138914}. 

\subsection{Interpreting the correlation patterns}

The positive awareness–workload coupling under depth-agnostic support aligns with reports that adding voice/text overlays without metric grounding limits the potential of LLMs for EFR suport \cite{GutirrezMaquiln2024}. Conversely, the negative association between confidence and workload under depth augmentation mirrors the promise of hybrid spatial language systems showing that when cues are metrically grounded and bound to detected objetcs (e.g., YOLOv8 labels with distances), users can rely on concise, high-value facts rather than reconstructing geometry from monocular hints \cite{YOLOv8,SAPKOTA2026103575,doi:10.1177/02783649251351658}. This pattern also aligns with reports of situational awareness that warn against overloading operators with uncalibrated cues \cite{doi:10.1518/001872095779049543} and with policy and operations guidance lacking sensor-backed support for EFR \cite{JRC138914}.

\section{Limitations}
This study was conducted with a relatively small sample size in a controlled, simulated environment, which  limits the generalizability of the findings. Although the distance-estimation task effectively captured key aspects of spatial reasoning relevant to first-responder operations, it could not fully replicate the complexity, stress, and time pressure inherent to real-world emergency contexts. The absence of dynamic elements such as moving agents, occlusions, or unpredictable hazards further constrains ecological validity.

Additionally, the present study assessed confidence and workload as indirect indicators of trust and cognitive efficiency. While these measures provided valuable insight into participants’ subjective experiences, they do not capture how trust and reliance on multimodal LLMs might evolve over time or under repeated exposure. Future research should therefore incorporate longitudinal designs to examine how users’ trust calibration and situational awareness develop as they gain familiarity with the system and encounter varying levels of automation reliability.

Several participants criticized the perceived latency of Qwen2.5vl:32b, despite being briefed that replies would arrive in about 30 seconds, due to hardware constraints (NVIDIA RTX 4080, 64 GB RAM, and an i9-13900K (3.00 GHz)). Participants described the wait as disruptive to situational flow and confidence in the assistant. Thus deceasing the waiting time is an important aspect for future work e.g. through unified Accelerated Processing Units, forthcoming GPUs, and tighter end-to-end information flow pipelining.

\section{Future Work}
Building on these findings, future studies should extend the current approach to real world field experiments that more closely approximate operational conditions faced by first responders. Scenarios including dynamic agents, partial visibility, and time-critical decisions would allow for a more realistic evaluation of multimodal LLM support for EFR under stress.


Moreover, integrating adaptive dialogue strategies, in which the LLM modulates the amount, timing, and modality of feedback based on estimated workload or situational demand, may further optimize the balance between situational awareness and cognitive load. Such adaptive interaction mechanisms, combined with depth-augmented spatial reasoning, have the potential to create robust, context-aware assistants that can flexibly support human operators in high-risk and/or rapidly evolving situations. Ultimately, these developments could pave the way for more trustworthy and resilient human–AI collaboration frameworks, where the system not only conveys spatial understanding efficiently but also aligns its communication dynamically with the human’s cognitive state and task context.

\section{Conclusion}
In this work we explore whether augmenting a vision language models with ground-truth depth information from a robot-mounted camera improves spatial context understanding for EFR. 

We developed and evaluated a prototype that fuses robot-mounted depth sensing and YOLOv8 detection with a vision language model (VLM) capable of verbalizing metrically-grounded distances of detected objects. 
In a mixed-reality toxic-smoke scenario, participants estimated distances to a victim and an exit window under three conditions: video-only, depth-agnostic VLM, and depth-augmented VLM. 

Depth-augmentation improved distance estimation, as we could observe lower distance error and variance than in the other conditions (human-only video and depth-agnostic LLM). Depth-augmentation improved situational awareness without increasing workload. Overall, depth-augmented VLM support is preferable to both human-only video inspection and depth-agnostic LLM commentary for distance-sensitive XR-EFR tasks. 

Situational awareness is a core competence for emergency first response (EFR), especially in complex and time-critical scenarios. In this work we demonstrate how human augmentation through robotic systems can increase situational awareness, improving human spatial reasoning and support decision-relevant judgments under time pressure.

\bibliographystyle{ACM-Reference-Format}
\bibliography{AHS26}

\appendix

\end{document}